\documentclass[prd,showpacs,showkeys,nofootinbib,floatfix,
               fleqn,preprint,12pt,tightenlines]{revtex4-1} 

\usepackage{amsfonts,amsmath,amssymb}
\usepackage{bm}
\usepackage{verbatim}
\usepackage{graphicx}

\newcommand{\beq}{\begin{equation}}
\newcommand{\eeq}{\end{equation}}
\newcommand{\beqa}{\begin{eqnarray}}
\newcommand{\eeqa}{\end{eqnarray}}
\newcommand{\bsubeqs}{\begin{subequations}}
\newcommand{\esubeqs}{\end{subequations}}

\begin{document}
\begin{widetext}
%
%
\noindent LHEP  \textbf{2022}, 312 (2022)  \hfill    arXiv:2207.03453
%
%
\newline\vspace*{3mm}
\end{widetext}

\title{Q-field from a 4D-brane:
Cosmological constant cancellation and Minkowski attractor}

\author{\vspace*{5mm} F.R. Klinkhamer}
\email{frans.klinkhamer@kit.edu}
\affiliation{Institute for Theoretical Physics,
Karlsruhe Institute of Technology (KIT),\\
76128 Karlsruhe,  Germany\\ \\}


\begin{abstract}
\vspace*{5mm}\noindent
A 4D-brane realization of $q$-theory has been proposed a few years ago.
The present paper studies the corresponding late-time cosmology and
establishes the dynamic cancellation of an initial cosmological constant
and an attractor behavior towards Minkowski spacetime.
\end{abstract}

\maketitle

\section{Introduction}
\label{sec:Introduction}

The cosmological constant problem is perhaps the most important
problem of modern physics~\cite{Weinberg1989}.
A condensed-matter-inspired approach has been proposed
and goes under the name of
$q$-theory~\cite{KlinkhamerVolovik2008a,KlinkhamerVolovik2008b,                      KlinkhamerVolovik2010}.
A particular realization of $q$-theory takes its cue from the physics
of a freely suspended two-dimensional material film~\cite{KatsLebedev2015}
and uses a generalization to four-dimensional ``films''
or 4D-branes~\cite{KlinkhamerVolovik-JETPL-2016-4Dbrane}.

This 4D-brane realization of $q$-theory has a dimensionless chemical
potential $\mu$ (details will be given shortly) and may be of relevance
to a recent suggestion to replace the big bang by a quantum phase
transition~\cite{KlinkhamerVolovik2022-BBasTopQPT}.

The goal of the present paper is to discuss the corresponding late-time
cosmology, where the asymptotic vacuum energy density may or may not vanish.
In particular, we would like to establish a possible attractor behavior
towards Minkowski spacetime with a vanishing vacuum energy density.
Throughout, we use natural units with $\hbar=1$ and $c=1$.\vspace*{-5mm}

\section{Action and field equations}
\label{sec:Action-field-equations}

The following 4D-brane action has been proposed in
Ref.~\cite{KlinkhamerVolovik-JETPL-2016-4Dbrane}:\vspace*{-1mm}
\begin{eqnarray}
\label{eq:Action-4Dbrane}
\hspace*{-0mm}
S &=&- \int
d^4x\, \sqrt{-g}\,\left[
\frac{R}{16\pi G_{N}} +
\epsilon\left(\frac{n}{\sqrt{-g}}\right)+
\mathcal{L}^{M}[\psi]\right]
+ \mu \int\,d^4x \;n \,,
\end{eqnarray}
with a Lorentzian signature ($-,\,+,\,+,\,+$)
of the metric $g_{\alpha\beta}$, so that
its determinant $g \equiv \det(g_{\alpha\beta})$ is negative.
In \eqref{eq:Action-4Dbrane},
the term $\mathcal{L}^{M}(\psi)$ is the matter Lagrange density
for a generic matter field $\psi$ and
the potential $\epsilon(x)$ is an
essentially arbitrary function of $x$
(the stability of the equilibrium state gives some conditions
on the potential~\cite{KlinkhamerVolovik2008a}).
The crucial new ingredient of the above action is
$n$ as the 4D analog of the particle density of a 2D
membrane. This 4D density $n$ may refer to  ``spacetime atoms,''
with the corresponding nonzero chemical potential $\mu$
(further discussion will be given in Sec.~\ref{sec:Outlook}).

At this moment, it may be helpful
to explain the terminology ``spacetime atoms.''
The physics of the freely suspended two-dimensional material film
has been reviewed in the first two paragraphs
of Sec.~3 in Ref.~\cite{KlinkhamerVolovik-JETPL-2016-4Dbrane},
with mention of the original reference~\cite{KatsLebedev2015}.
The film there is made out of atoms, whose basic structure is known
to be described by the Schr\"{o}dinger equation.
The Hamiltonian of this two-dimensional material film
has been generalized to a four-dimensional action and, for this reason, we
can perhaps speak of a system describing ``spacetime atoms,''
whose structure is, of course, completely unknown at the present moment.
What matters, for the following, is the structure of the
action \eqref{eq:Action-4Dbrane} with an unknown (conserved)
quantity $n$ having mass dimension 4.

The resulting $n$-field equation from \eqref{eq:Action-4Dbrane}
reads\vspace*{-1mm}
\begin{subequations}
\label{eq:variation-4D-brane-mu-q}
\begin{eqnarray}
\label{eq:variation-4D-brane-mu}
\frac{d\epsilon}{dq}&=& \mu \,,
\\[0mm]
\label{eq:variation-4D-brane-q}
q                   &\equiv& \frac{n}{\sqrt{-g}}\,.
\end{eqnarray}
\end{subequations}
Here, we have identified a scalar $q$-field with
mass dimension 4, so that $\mu$ has mass dimension 0
and this $q$-field realization may be relevant to
the big-bang discussion of Ref.~\cite{KlinkhamerVolovik2022-BBasTopQPT}.

The resulting gravitational equation from \eqref{eq:Action-4Dbrane} is the
standard Einstein equation,
\beq
\label{eq:Einstein-eq-4D-brane}
\frac{1}{8\pi G_{N}}
\left( R_{\alpha\beta}-\frac{1}{2}\,R\,g_{\alpha\beta}\right)=
\rho_{V}(q)\, g_{\alpha\beta}+T^{M}_{\alpha\beta}\,,
\eeq
with the following gravitating vacuum energy density:
\beq
\label{eq:rho-V}
\rho_{V}(q)
=  \epsilon(q)  - q\,\frac{d\epsilon(q)}{d q}
=  \epsilon(q)  - \mu\, q\,,
\eeq
where \eqref{eq:variation-4D-brane-mu} has been used to get
the final expression for $\rho_{V}(q)$. Note that
the constant $\mu$ here traces back to the 4D-brane
action \eqref{eq:Action-4Dbrane}.
This situation is different from the one for
the $q$-field in the 4-form realization, where
$\mu$ appears as an integration constant of the
solution~\cite{KlinkhamerVolovik2008a,KlinkhamerVolovik2008b}.

In terms of the gravitating vacuum energy density \eqref{eq:rho-V},
we can rewrite the $n$-field equation \eqref{eq:variation-4D-brane-mu} as
\beq
\label{eq:rho-V-conservation}
\frac{d \rho_{V}(q)}{d q}=0\,.
\eeq
Remark that for a homogeneous $q$-field [i.e., $q=q(t)$
in a suitable coordinate system],
we can multiply \eqref{eq:rho-V-conservation} by $d q/d t$ to get
\beq
\label{eq:rho-V-time conservation}
\frac{d \rho_{V}(t)}{d t}=0\,,
\eeq
which carries over to cosmology, as will be discussed
in the next section.

\section{Cosmology: Minkowski attractor}
\label{sec:Cosmology}

\subsection{Setup}
\label{subsec:Setup}

For cosmology, we take
the standard spatially flat Robertson--Walker (RW) metric
with cosmic scale factor $a(t)$ and Hubble parameter
$H\equiv [da/dt]/a$. We also
add a homogeneous perfect fluid for the matter component
with a constant equation-of-state parameter
\mbox{$w_M \equiv  P_M/ \rho_M$.}

Next, introduce dimensionless variables:
the cosmic time coordinate $\tau$, the Hubble parameter
$h\equiv a^{-1}\,da/d\tau$,
the matter energy density $r_{M}$,
and the vacuum energy density $r_{V}$.
Then,  the dimensionless ordinary differential equations (ODEs)
are as follows:
\bsubeqs\label{eq:4Dbrane-dimensionless-ODEs}
\beqa
\label{eq:4Dbrane-dimensionless-ODE-rMdoteq}
\hspace*{0mm}
&&
\dot{r}_{M} +  3\, h\,\big( 1 + w_M  \big) \,r_{M} = +\Gamma\,,
\\[2mm]
\label{eq:4Dbrane-dimensionless-ODE-rVdoteq}
\hspace*{0mm}
&&
\dot{r}_{V} = -\Gamma\,,
\\[2mm]
\label{eq:4Dbrane-dimensionless-ODE-1stF}
\hspace*{0mm}
&&
3\,h^2  = 8\,\pi \, \big(r_{M}+r_\text{V} \big)\,,
\\[2mm]
\label{eq:4Dbrane-dimensionless-ODE-2ndF}
\hspace*{0mm}
&&
2\,\dot{h}+3\,h^2 =- 8\,\pi \, \big( w_M\,r_{M}-  r_\text{V} \big) \,,
\eeqa
\esubeqs
where the overdot stands for
differentiation with respect to the dimensionless cosmic
time coordinate $\tau$
and the source term $\Gamma$ models
vacuum-matter energy exchange, as discussed by
Ref.~\cite{KlinkhamerSavelainenVolovik2016} in general terms.

\subsection{General $\boldsymbol{\rho_{V}}$: Analytic solution}
\label{subsec:General-case}

An explicit calculation of $\Gamma$ entering the
ODEs \eqref{eq:4Dbrane-dimensionless-ODEs}
was presented in Ref.~\cite{KlinkhamerVolovik-MPLA-2016}
and gave a Zeldovich--Starobinsky-type~\cite{ZeldovichStarobinsky1977}
source term:
\beq
\label{eq:Gamma-particle-production}
\Gamma_\text{particle-production}
= \frac{1}{36}\,\gamma\,|h|\,\mathcal{R}^2
= \gamma\,|h|\,\left( \dot{h}+2\,h^2 \right)^2\,,
\eeq
with $\mathcal{R}$ being the Ricci curvature scalar
in terms of dimensionless variables.
Observe that equations \eqref{eq:4Dbrane-dimensionless-ODE-rMdoteq}
and \eqref{eq:4Dbrane-dimensionless-ODE-rVdoteq}
are time-reversal noninvariant for
the source term as given by  \eqref{eq:Gamma-particle-production}.
This time-reversal noninvariance corresponds to a dissipative effect,
actually a quantum-dissipative effect as particle creation is a
true quantum phenomenon.

The ODEs \eqref{eq:4Dbrane-dimensionless-ODEs},
for the source term \eqref{eq:Gamma-particle-production} and
$w_M=1/3$, are \emph{exactly the same}
as those in Ref.~\cite{KlinkhamerVolovik-MPLA-2016},
which were derived with the 4-form realization of the
$q$-field.\footnote{The 4-form theory considered in
Refs.~\cite{KlinkhamerVolovik2008a,KlinkhamerVolovik2008b,%
KlinkhamerVolovik2010,KlinkhamerVolovik-MPLA-2016} is purely four-dimensional,
so that the values of $q$ are continuous and not quantized
as happens for other types of 4-form theories~\cite{BoussoPolchinski2000}.}
Hence, there is the \emph{same} attractor behavior towards
Minkowski spacetime; see, in Ref.~\cite{KlinkhamerVolovik-MPLA-2016},
the numerical results of Fig.~3 and the analytic solution of
App.~B.

It may be instructive to recall the main steps for
getting this exact solution. For the special case
of relativistic matter, $w_M=1/3$, we can add the
two Friedmann equations \eqref{eq:4Dbrane-dimensionless-ODE-1stF}
and \eqref{eq:4Dbrane-dimensionless-ODE-2ndF}, in order to eliminate
$r_M$.
The resulting equation relates the combination
$\dot{h}+2\,h^2$ to $r_{V}$.
The special choice \eqref{eq:Gamma-particle-production}
for $\Gamma$ involves the very same combination $\dot{h}+2\,h^2$
and \eqref{eq:4Dbrane-dimensionless-ODE-rVdoteq} reduces to
a single ODE for $r_{V}(\tau)$:
\beq
\label{eq:4Dbrane-dimensionless-ODE-rVdoteq-reduced}
\frac{1}{|h|}\,\frac{d r_{V} }{d \tau}
= -\gamma\,\Big[ (16\pi/3)\, r_{V} \Big]^2\,.
\eeq
Changing the $\tau$ coordinate to
\beq
\label{eq:chi-def}
\chi = \ln|a(\tau)|\,,
\eeq
we obtain the following ODE:
\beq
\label{eq:4Dbrane-dimensionless-ODE-rVdoteq-reduced-chi}
\frac{d r_{V}}{d \chi}
= -(16\pi/3)^2\,\gamma\;\big[  r_{V} \big]^2\,.
\eeq
The solution (denoted by a bar) is
\beq
\label{eq:rV-chi-solution}
\overline{r}_{V}(\chi)=
\frac{1}{(16\pi/3)^2\;\gamma\;\left(\chi -\chi_{0} \right)}\,,
\eeq
with an integration constant $\chi_{0}$. The corresponding solution
$\overline{h}(\chi)$ involves the exponential integral function
``$\text{Ei}(\chi)$'' and is given
in App.~B of Ref.~\cite{KlinkhamerVolovik-MPLA-2016}.
As $a(\tau)$ goes to infinity with $\tau\to\infty$,
so does $\chi$ and the vacuum energy density $r_{V}$
from the exact solution \eqref{eq:rV-chi-solution}
is seen to drop to zero.

\subsection{Specific $\boldsymbol{\rho_{V}(q)}$: ODEs}
\label{subsec:Specific-case}

There is, however, an important caveat
for the general discussion in Sec.~\ref{subsec:General-case}:
the dimensionless vacuum energy density $r_{V}$
corresponding to the dimensional quantity
$\rho_{V}$ from \eqref{eq:rho-V} should be
able to reach the value zero.

Denote the dimensionless version of the cosmological constant
by $\lambda$
and the dimensionless version of the
$q$-variable from \eqref{eq:variation-4D-brane-q} by $\xi$
(the chemical potential $\mu$ is already dimensionless).
Then, we can split the dimensionless version of the
energy density $\epsilon$ from the action \eqref{eq:Action-4Dbrane}
into a constant part $\lambda$
and a nonconstant part $\widetilde{\epsilon}$,
\beq\label{eq:epsilontilde}
\epsilon[\xi] \equiv \lambda + \widetilde{\epsilon}\,[\xi]\,,
\eeq
and require for the existence of a Minkowski attractor:
\beq
\label{eq:rV-condition-xibar}
\exists\, \overline{\xi} \in \mathbb{R}^{+}: \;\;
r_{V}[\, \overline{\xi}\, ]=
\lambda + \widetilde{\epsilon}\left[\, \overline{\xi}\, \right]
- \mu\,\overline{\xi} =0\,.
\eeq
The issue, now, is whether or not there exists such a
$\overline{\xi}$ and, if there exists such a $\overline{\xi}$,
whether or not it can be reached in the late-time cosmology.

For a specific realization of the vacuum energy density
$r_{V}$ as a function of $\xi$
and with the Hubble parameter $h\equiv \dot{a}/a$,
the ODEs \eqref{eq:4Dbrane-dimensionless-ODEs}
take the following form:
\bsubeqs\label{eq:4Dbrane-dimensionless-xi-h-ODEs}
\beqa
\label{eq:4Dbrane-dimensionless-xi-h-ODE-rMdoteq}
\hspace*{0mm}
&&\dot{r}_{M} +  3\, h\,\big( 1 + w_M \big) \,r_{M}
= +\Gamma\,,
\\[2mm]
\label{eq:4Dbrane-dimensionless-xi-h-ODE-xidoteq}
\hspace*{0mm}
&&\dot{\xi}\; \frac{d r_{V}[\xi]}{d \xi}
= -\Gamma\,,
\\[2mm]
\label{eq:4Dbrane-dimensionless-xi-h-ODE-1stF}
\hspace*{0mm}
&&3\,h^2
 =
8\,\pi \, \Big(r_{M}+r_\text{V}[\xi] \Big)\,,
\\[2mm]
\label{eq:4Dbrane-dimensionless-xi-h-ODE-2ndF}
\hspace*{0mm}
&&2\,\dot{h}+3\,h^2
 =
- 8\,\pi \, \Big( w_M\,r_{M}-  r_\text{V}[\xi] \Big) \,,
\\[2mm]
\label{eq:4Dbrane-dimensionless-xi-h-ODE-rV}
\hspace*{0mm}
&&r_{V}[\xi] = \lambda + \widetilde{\epsilon}\,[\xi]- \mu\,\xi \,,
\eeqa
\esubeqs
where the dependence of $r_{V}$
on the vacuum variable $\xi$ has been made explicit.

As a simple example, take
\bsubeqs\label{eq:epsilontilde-Ansatz-lambda-range-xibar}
\beq\label{eq:epsilontilde-Ansatz}
\widetilde{\epsilon}\,[\xi]= \xi^2 \,,
\eeq
which turns condition \eqref{eq:rV-condition-xibar}
into a quadratic for $\overline{\xi}$.
For a given positive value of $\mu$,
there is now a suitable value $\overline{\xi}$
for \emph{any} value of the dimensionless cosmological constant
in the following range:
\beq
\label{eq:epsilontilde-Ansatz-lambda-range}
\lambda \leq \mu^2/4\,,
\eeq
where one possible value of $\overline{\xi}$ is given by
\beq
\label{eq:epsilontilde-Ansatz-lambda-range-xibarsol}
\overline{\xi}= \frac{1}{2}\,\left(\mu + \sqrt{\mu^2-4\,\lambda}\right)\,.
\eeq
\esubeqs
For $0<\lambda < \mu^2/4$, the other possible value of $\overline{\xi}$
has a minus sign in front of the square root
on the right-hand side of \eqref{eq:epsilontilde-Ansatz-lambda-range-xibarsol}.

A further point is the choice of $\Gamma$ so that
the numerics works in the simplest way.
A suitable \textit{ad hoc} choice is
\beq
\label{eq:Gamma-ad-hoc}
\Gamma_\text{ad-hoc}= \widetilde{\gamma}\,|h|\,h^4\,.
\eeq
This basically has the structure of
expression \eqref{eq:Gamma-particle-production},
but the squared Ricci factor has been simplified to
the fourth power of the Hubble parameter $h$.
Observe, again, that the
ODE \eqref{eq:4Dbrane-dimensionless-xi-h-ODE-xidoteq}
with source term \eqref{eq:Gamma-ad-hoc} is time-reversal noninvariant.

\subsection{Specific $\boldsymbol{\rho_{V}(q)}$: Numerical results}
\label{subsec:Numerical-results}

\begin{figure}[t]
\vspace*{-5mm}
\begin{center}
\hspace*{-9mm}
\includegraphics[width=1.1\textwidth]{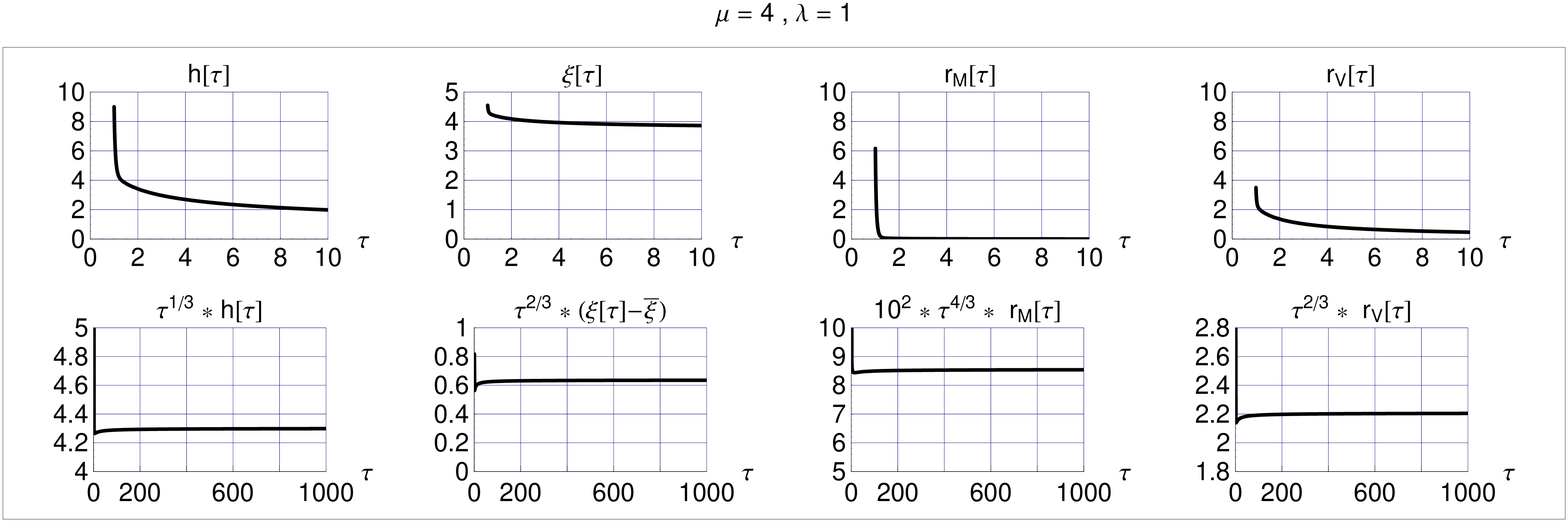}
\end{center}\vspace*{-4mm}
\caption{Numerical solution of the
ODEs \eqref{eq:4Dbrane-dimensionless-xi-h-ODEs} with
source term \eqref{eq:Gamma-ad-hoc}
and energy density \eqref{eq:epsilontilde-Ansatz},
for parameters
$w_M=1/3$, $\mu=4$, $\lambda=1$, and $\widetilde{\gamma}=10^{-3}$.
The boundary conditions at $\tau=\tau_\text{bcs}=1$
are $\{ h,\,\xi,\,r_{M}\}= \{   9, 4.55,
6.16616 \}$, 
where the $r_{M}$ value has been obtained from
the first Friedman equation \eqref{eq:4Dbrane-dimensionless-xi-h-ODE-1stF}.
The top row shows the three basic variables:
the Hubble parameter $h=\dot{a}/a$,
the dimensionless vacuum variable $\xi$,
and the matter energy density $r_{M}$.
The bottom row shows their asymptotic behavior:
$h(\tau) \sim \tau^{-1/3}$,
$\xi(\tau)-\overline{\xi} \sim \tau^{-2/3}$,
$r_M(\tau) \sim \tau^{-4/3}$,
and $r_{V}(\tau) \sim \tau^{-2/3}$, where
$\overline{\xi}$ is
given by \eqref{eq:epsilontilde-Ansatz-lambda-range-xibarsol}.
Both rows show the corresponding vacuum energy density $r_{V}$
from \eqref{eq:4Dbrane-dimensionless-xi-h-ODE-rV}
and \eqref{eq:epsilontilde-Ansatz}.
}
\label{fig:num-sol-lambdaplus1-4brane}
\vspace*{-4mm}
%
\begin{center}
\hspace*{-9mm}  
\includegraphics[width=1.1\textwidth]{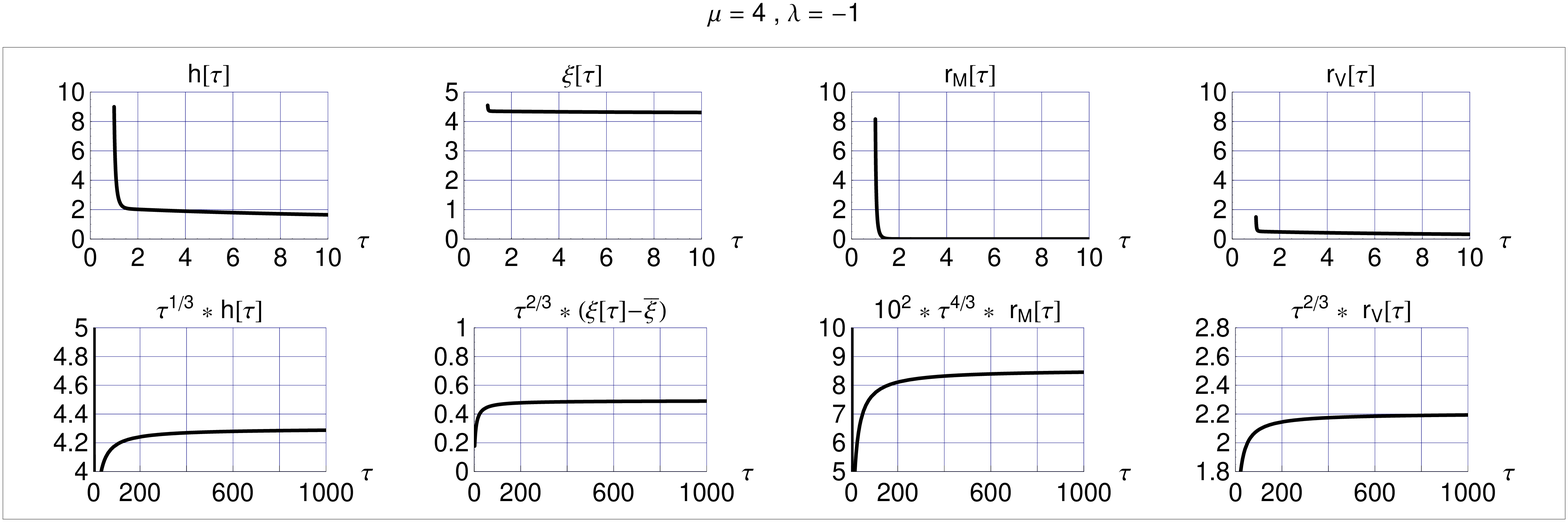}
\end{center}
\vspace*{-4mm}
\caption{Numerical solution of the
ODEs \eqref{eq:4Dbrane-dimensionless-xi-h-ODEs} with
source term \eqref{eq:Gamma-ad-hoc}
and energy density \eqref{eq:epsilontilde-Ansatz},
for parameters
$w_M=1/3$, $\mu=4$, $\lambda=-1$, and $\widetilde{\gamma}=10^{-3}$.
The boundary conditions at $\tau=\tau_\text{bcs}=1$
are $\{ h,\,\xi,\,r_{M}\}= \{   9, 4.55,
8.16616\}$, 
where the $r_{M}$ value has been obtained from
the first Friedman equation \eqref{eq:4Dbrane-dimensionless-xi-h-ODE-1stF}.
The variables shown are explained in the
caption of Fig.~\ref{fig:num-sol-lambdaplus1-4brane}.
}
\label{fig:num-sol-lambdaminus1-4brane}
\vspace*{0mm}
\end{figure}

Numerical results from the ODEs \eqref{eq:4Dbrane-dimensionless-xi-h-ODEs},
with \eqref{eq:epsilontilde-Ansatz-lambda-range-xibar}
and \eqref{eq:Gamma-ad-hoc},
are presented in Figs.~\ref{fig:num-sol-lambdaplus1-4brane}
and \ref{fig:num-sol-lambdaminus1-4brane}
for $\mu=4$ and $\lambda=\pm1$
(similar results have been obtained for $\mu=4$ and $\lambda=0$).
The asymptotic behavior in both figures is essentially the same,
\beq
\label{eq:h-xi-asymptotics}
h(\tau) \sim \big(\widetilde{\gamma}\,\tau\big)^{-1/3}\,,
\quad
r_{V}(\tau) \sim \big(\widetilde{\gamma}\,\tau\big)^{-2/3}\,.
\eeq
This asymptotic behavior also follows directly from the ODEs
\eqref{eq:4Dbrane-dimensionless-ODE-rVdoteq}
and \eqref{eq:4Dbrane-dimensionless-ODE-1stF}, for
the source term $\Gamma$ from \eqref{eq:Gamma-ad-hoc}
and assuming that $r_M$ is negligible compared to $r_{V}$.
The asymptotic values of $\xi(\tau)$ in
Figs.~\ref{fig:num-sol-lambdaplus1-4brane}
and \ref{fig:num-sol-lambdaminus1-4brane} are numerically close
to the analytic results
from \eqref{eq:epsilontilde-Ansatz-lambda-range-xibarsol}.

The asymptotic behavior \eqref{eq:h-xi-asymptotics}
illustrates the Minkowski attractor behavior ($r_{V} \to 0$).
Indeed, we find numerically the same attractor behavior
at the following four corners of the rectangle $\Delta$
of initial conditions:
\beq
\label{eq:corners}
\{ h(\tau_\text{bcs}),\xi(\tau_\text{bcs})\}=
\{   9 \pm 1/10, 455 /100\pm 1/100\}\,,
\eeq
equally for $\lambda=1$ and  $\lambda=-1$, at $\mu=4$.
We have also established numerically the same attractor behavior
for several random points over the rectangle $\Delta$.

These numerical results suggest that, for $\mu=4$ and $|\lambda|\leq 1$,
the attractor domain $\mathcal{D}$ in the plane $\mathbb{R}^2$ of
initial conditions is finite and
includes the above-mentioned rectangle,
\beq
\label{eq:attractor-domain-D}
\mathcal{D}^{(\mu=4,\,|\lambda|\leq 1)} \supseteq \Delta =
\Big\{
\{ h(\tau_\text{bcs}),\xi(\tau_\text{bcs})\}\;\Big|\;
8.9 \leq h(\tau_\text{bcs}) \leq 9.1
\;\wedge\;
4.54 \leq \xi(\tau_\text{bcs}) \leq 4.56 \Big\}\,.
\eeq
The actual attractor domain $\mathcal{D}$ can be expected to be
larger than the rectangle indicated.

\begin{figure}[t]
\vspace*{-5mm}
\begin{center}
\hspace*{-9mm}
\includegraphics[width=1.1\textwidth]{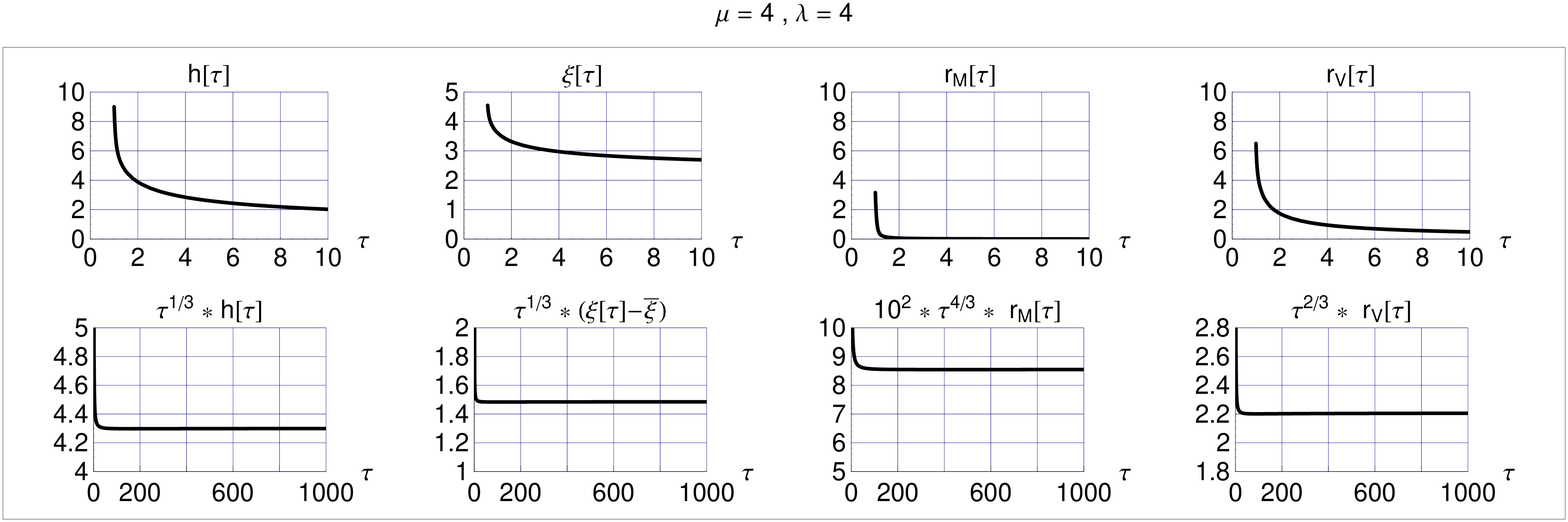}
\end{center}\vspace*{-4mm}
\caption{Numerical solution of the
ODEs \eqref{eq:4Dbrane-dimensionless-xi-h-ODEs} with
source term \eqref{eq:Gamma-ad-hoc}
and energy density \eqref{eq:epsilontilde-Ansatz},
for parameters
$w_M=1/3$, $\mu=4$, $\lambda=4$, and $\widetilde{\gamma}=10^{-3}$.
The boundary conditions at $\tau=\tau_\text{bcs}=1$
are $\{ h,\,\xi,\,r_{M}\}= \{   9, 4.55,
3.16616 \}$, 
where the $r_{M}$ value has been obtained from
the first Friedman equation \eqref{eq:4Dbrane-dimensionless-xi-h-ODE-1stF}.
The variables shown are explained in the
caption of Fig.~\ref{fig:num-sol-lambdaplus1-4brane}, but
the asymptotic behavior of the dimensionless vacuum variable is now
different: $\xi(\tau)-\overline{\xi} \sim \tau^{-1/3}$.}
\label{fig:num-sol-lambdaplus4-4brane}
\vspace*{0mm}
\end{figure}

In addition, we have similar numerical results for $\mu=4$ and $\lambda=4$
(shown in Fig.~\ref{fig:num-sol-lambdaplus4-4brane}
and Table~\ref{tab:num-sol-mu4-lambda4}),
which will be discussed further in Sec.~\ref{subsec:Discussion}.

\subsection{Discussion}
\label{subsec:Discussion}

Following-up on the attractor-domain discussion of
the last subsection with numerical results, we emphasize that
the analytic solution~\cite{KlinkhamerVolovik-MPLA-2016}
is clear about having a finite attractor domain. Let us give
the details (expanding on the statement from the last paragraph
of App.~B in that reference):
$\xi(\tau_\text{bcs})$ must be
such as to make $r_{V}(\tau_\text{bcs})$ nonnegative
and $h(\tau_\text{bcs})$ must also be nonnegative
with a further condition
that traces back to \eqref{eq:4Dbrane-dimensionless-xi-h-ODE-1stF}.
These conditions can be summarized as follows:
\beq
\label{eq:h-rV-conditions}
h(\tau_\text{bcs})
\geq
\sqrt{\frac{8\pi}{3}\,r_{V}(\tau_\text{bcs})}
\geq 0 \,.
\eeq
How the actual attractor domain looks
in the $\{ h(\tau_\text{bcs}),\xi(\tau_\text{bcs})\}$ plane
depends on the details of the \textit{Ansatz} for $\widetilde{\epsilon}\,[\xi]$
and the numerical values of $\mu$ and $\lambda$,
possibly obeying a condition similar
to \eqref{eq:epsilontilde-Ansatz-lambda-range}.
Incidentally, there is no such condition on $\lambda$,
for given $\mu$, if the $\widetilde{\epsilon}$
\textit{Ansatz} is changed to, for example,
$\widetilde{\epsilon}\,[\xi]=(\xi^2)^{1/4} + (\xi^2)^{-1/4}$,
which allows for the cancellation of \emph{any} value of
$\lambda\in \mathbb{R}$ for arbitrary $\mu$.

There is, however, a puzzle. Namely,
the numerical results for $\mu=4$ and $\lambda=\pm1$
in Figs.~\ref{fig:num-sol-lambdaplus1-4brane}
and \ref{fig:num-sol-lambdaminus1-4brane}
show that Minkowski spacetime is approached,
but how can that be as $\mu=4$  differs from the
fine-tuned value $\mu_{0}$ for the Minkowski vacuum of
Ref.~\cite{KlinkhamerVolovik2008a}? Incidentally,
for the $\widetilde{\epsilon}(\xi)$ \textit{Ansatz} in
\eqref{eq:epsilontilde-Ansatz}, the equilibrium value
of the chemical potential is given by $\mu_{0}=2\,\sqrt{\lambda}$,
provided $\lambda$ is nonnegative
(all the more surprising that our numerical
solution can approach Minkowski spacetime also for negative
$\lambda$!).

The answer is simply that the Minkowski vacuum of
Ref.~\cite{KlinkhamerVolovik2008a} holds for \emph{static} fields,
whereas our numerical solution is \emph{nonstatic},
$\xi=\xi(\tau)$ with $\dot{\xi} \ne 0$.

The numerical results for $\mu=4$ and $\lambda=\pm1$
do not appear to have a rigorous ``limit'' as $\tau\to\infty$
and the fields are essentially time-dependent.
This is reminiscent of Dolgov's model~\cite{Dolgov1985,Dolgov1997}
(see also Fig.~2~ in  Ref.~\cite{KlinkhamerVolovik2010}
and App.~A in Ref.~\cite{EmelyanovKlinkhamer2012}),
even though the time-dependence of $h=h(\tau)$ and $\xi=\xi(\tau)$
in our 4D-brane model \emph{diminishes} with time,
whereas Dolgov's time-dependence stays constant with time
(specifically, a linear time dependence of the massless vector field).
Still, the final state with $q=\overline{q}$ is, in general, not
the static-equilibrium state with $q=q_{0}$ and the question
remains whether or not such a situation is acceptable
(especially as concerns
the gravitational dynamics of a solar-system-type subsystem;
see Sec.~I of Ref.~\cite{EmelyanovKlinkhamer2012} for further
discussion and references).

For the special case $\mu=4$ and $\lambda=4$, there may be a
strict limit, as we then reach the genuine Minkowski vacuum of
Ref.~\cite{KlinkhamerVolovik2008a} at $\mu_{0}=4$,
with $\xi_{0}=2$ from \eqref{eq:epsilontilde-Ansatz-lambda-range-xibarsol}.
The $\xi(\tau)$ values in Table~\ref{tab:num-sol-mu4-lambda4}
appear to approach the equilibrium value $\xi_{0}=2$.
The corresponding dimensional vacuum variable $q(t)$ then
approaches the equilibrium value $q_{0}$, around which the
vacuum energy density has a quadratic behavior,
$\rho_{V}(q) \sim c_{2}\, \big(q-q_{0}\big)^{2}$
with a positive constant $c_{2}$ of mass dimension $-4$
(here, $c_{2}\propto E_\text{Planck}^{-4}$).\vspace*{-2mm}

\begin{table*}[t]
\vspace*{-0mm}
\caption{Function values from the numerical solution of
Fig.~\ref{fig:num-sol-lambdaplus4-4brane},
showing only 3 significant digits.
\vspace*{2mm}}
\label{tab:num-sol-mu4-lambda4}
\begin{ruledtabular}
\renewcommand{\tabcolsep}{-0pc}       
\renewcommand{\arraystretch}{1.125}   
\begin{tabular}{ccccc}
$\;\;\tau$ & $h(\tau)$ & $\xi(\tau)$ & $r_{M}(\tau)$ & $r_{V}(\tau)\;\;$ \\
\hline
$\;\;10^{2}$ & $0.926  $&$2.32 $&$1.85\times{10}^{-4} $&$0.102\;\;$\\
$\;\;10^{3}$ & $0.430  $&$2.15 $&$8.55\times{10}^{-6} $&$0.0221\;\;$\\
$\;\;10^{4}$ & $0.200  $&$2.07 $&$3.97\times{10}^{-7} $&$0.00476\;\;$\\
$\;\;10^{5}$ & $0.0927 $&$2.03 $&$1.84\times{10}^{-8} $&$0.00102\;\;$\\
$\;\;10^{6}$ & $0.0430 $&$2.01 $&$8.56\times{10}^{-10}$&$0.000221\;\;$\\
\hline
\end{tabular}
\end{ruledtabular}
%
%
%
\vspace*{-2mm}
\end{table*}

\section{Outlook}
\label{sec:Outlook}

The present paper has studied the cosmological behavior of a $q$-field
in the 4D-brane realization~\cite{KlinkhamerVolovik-JETPL-2016-4Dbrane}.
The physical interpretation of this
$q$-field scalar relates to a four-dimensional number
density $n$ of ``spacetime atoms,'' with a corresponding chemical
potential $\mu$. This number density $n$ is a \emph{new} variable,
as the action \eqref{eq:Action-4Dbrane} makes clear.
It leads to an additional conservation
equation \eqref{eq:rho-V-conservation} of the corresponding vacuum energy
density $\rho_{V}(q)$.

A similar model has been presented
recently~\cite{Klinkhamer2022-ext-unimod-grav}, where there is
also a  four-dimensional number density $n$
with a corresponding chemical potential $\mu$.
But, in that case, there is no need for a new variable as $n$
is simply proportional to $\sqrt{-g}$,
in terms of the \emph{already available}
metric determinant $g \equiv \det (g_{\alpha\beta})$.
[This identification implies the restriction of the allowed
coordinate transformations to those with unit Jacobian.]
In that case, there is no additional conservation equation
for $\rho_{V}$. Yet, in a cosmological context,
the Friedmann equations do contain the equation $d\rho_{V}/d t=0$;
see the last paragraph of
Sec.~VI B in Ref.~\cite{Klinkhamer2022-ext-unimod-grav}.
Moreover, there appears an attractor behavior towards Minkowski spacetime
provided there is vacuum-matter energy exchange.

Both manifestations of $n$ apparently involve some fine-scale underlying
structure of spacetime, called ``atoms of spacetime'' for the 4D-brane
realization~\cite{KlinkhamerVolovik-JETPL-2016-4Dbrane}
and a ``spacetime crystal'' for the extended-unimodular-gravity
approach~\cite{Klinkhamer2022-ext-unimod-grav}.
The outstanding question is what the actual substructure of
spacetime really is. A partial answer can perhaps be
obtained from  nonperturbative superstring theory as
formulated by the IIB matrix
model~\cite{IKKT-1997,Aoki-etal-review-1999,Klinkhamer2021-master}.\vspace*{-2mm}

\begin{acknowledgments}
We thank G.E. Volovik and the referee for useful comments
on a first version of this paper
and we acknowledge support by the KIT-Publication Fund of
the Karlsruhe Institute of Technology.
\end{acknowledgments}

\end{document}